

\def\unlock{
 \catcode`@=11 }
\unlock
\def\lock{\catcode`@=12 }

\newskip\@@@@@
\newskip\@@@@@two
\newskip\t@mp


 \font\twentyfourrm=cmr10               scaled\magstep4
 \font\seventeenrm=cmr10                scaled\magstep3
 \font\fourteenrm=cmr10                 scaled\magstep2
 \font\twelverm=cmr10                   scaled\magstep1
 \font\ninerm=cmr9            \font\sixrm=cmr6
 \font\twentyfourbf=cmbx10              scaled\magstep4
 \font\seventeenbf=cmbx10               scaled\magstep3
 \font\fourteenbf=cmbx10                scaled\magstep2
 \font\twelvebf=cmbx10                  scaled\magstep1
 \font\ninebf=cmbx9            \font\sixbf=cmbx5
 \font\twentyfouri=cmmi10 scaled\magstep4  \skewchar\twentyfouri='177
 \font\seventeeni=cmmi10  scaled\magstep3  \skewchar\seventeeni='177
 \font\fourteeni=cmmi10   scaled\magstep2  \skewchar\fourteeni='177
 \font\twelvei=cmmi10     scaled\magstep1  \skewchar\twelvei='177
 \font\ninei=cmmi9                         \skewchar\ninei='177
 \font\sixi=cmmi6                          \skewchar\sixi='177
 \font\twentyfoursy=cmsy10 scaled\magstep4 \skewchar\twentyfoursy='60
 \font\seventeensy=cmsy10  scaled\magstep3 \skewchar\seventeensy='60
 \font\fourteensy=cmsy10   scaled\magstep2 \skewchar\fourteensy='60
 \font\twelvesy=cmsy10     scaled\magstep1 \skewchar\twelvesy='60
 \font\ninesy=cmsy9                        \skewchar\ninesy='60
 \font\sixsy=cmsy6                         \skewchar\sixsy='60
 \font\twentyfourex=cmex10  scaled\magstep4
 \font\seventeenex=cmex10   scaled\magstep3
 \font\fourteenex=cmex10    scaled\magstep2
 \font\twelveex=cmex10      scaled\magstep1
 \font\elevenex=cmex10      scaled\magstephalf
 \font\twentyfoursl=cmsl10  scaled\magstep4
 \font\seventeensl=cmsl10   scaled\magstep3
 \font\fourteensl=cmsl10    scaled\magstep2
 \font\twelvesl=cmsl10      scaled\magstep1
 \font\ninesl=cmsl9
 \font\twentyfourit=cmti10 scaled\magstep4
 
 \font\fourteenit=cmti10   scaled\magstep2
 \font\twelveit=cmti10     scaled\magstep1
 \font\twentyfourtt=cmtt10 scaled\magstep4
 \font\twelvett=cmtt10     scaled\magstep1
 \font\twentyfourcp=cmcsc10 scaled\magstep4
 \font\twelvecp=cmcsc10    scaled\magstep1
 \font\tencp=cmcsc10
 \font\hepbig=cmdunh10     scaled\magstep5
 \font\hep=cmdunh10        scaled\magstep0
 \newfam\cpfam
 \font\tenfib=cmr10  
 \newcount\f@ntkey            \f@ntkey=0
 \def\samef@nt{\relax \ifcase\f@ntkey \rm \or\oldstyle \or\or
          \or\it \or\sl \or\bf \or\tt \or\caps \fi }

 \def\twentyfourpoint{\relax
     \textfont0=\twentyfourrm           \scriptfont0=\seventeenrm
                                        \scriptscriptfont0=\fourteenrm
     \def\rm{\fam0 \twentyfourrm \f@ntkey=0 }\relax
     \textfont1=\twentyfouri            \scriptfont1=\seventeeni
                                        \scriptscriptfont1=\fourteeni
      \def\oldstyle{\fam1 \twentyfouri\f@ntkey=1 }\relax
     \textfont2=\twentyfoursy           \scriptfont2=\seventeensy
                                        \scriptscriptfont2=\fourteensy
     \textfont3=\twentyfourex           \scriptfont3=\seventeenex
                                        \scriptscriptfont3=\fourteenex
     \def\it{\fam\itfam \twentyfourit\f@ntkey=4 }\textfont\itfam=\twentyfourit
     \def\sl{\fam\slfam \twentyfoursl\f@ntkey=5 }\textfont\slfam=\twentyfoursl
                                        \scriptfont\slfam=\seventeensl
     \def\bf{\fam\bffam \twentyfourbf\f@ntkey=6 }\textfont\bffam=\twentyfourbf
                                        \scriptfont\bffam=\seventeenbf
                                        \scriptscriptfont\bffam=\fourteenbf
     \def\tt{\fam\ttfam \twentyfourtt \f@ntkey=7 }\textfont\ttfam=\twentyfourtt
     \def\caps{\fam\cpfam \twentyfourcp \f@ntkey=8 }
                                        \textfont\cpfam=\twentyfourcp
     \setbox\strutbox=\hbox{\vrule height 23pt depth 8pt width\z@}
     \samef@nt}
 \def\fourteenpoint{\relax
     \textfont0=\fourteenrm          \scriptfont0=\tenrm
     \scriptscriptfont0=\sevenrm
      \def\rm{\fam0 \fourteenrm \f@ntkey=0 }\relax
     \textfont1=\fourteeni           \scriptfont1=\teni
     \scriptscriptfont1=\seveni
      \def\oldstyle{\fam1 \fourteeni\f@ntkey=1 }\relax
     \textfont2=\fourteensy          \scriptfont2=\tensy
     \scriptscriptfont2=\sevensy
     \textfont3=\fourteenex          \scriptfont3=\twelveex
     \scriptscriptfont3=\tenex
     \def\it{\fam\itfam \fourteenit\f@ntkey=4 }\textfont\itfam=\fourteenit
     \def\sl{\fam\slfam \fourteensl\f@ntkey=5 }\textfont\slfam=\fourteensl
     \scriptfont\slfam=\tensl
     \def\bf{\fam\bffam \fourteenbf\f@ntkey=6 }\textfont\bffam=\fourteenbf
     \scriptfont\bffam=\tenbf     \scriptscriptfont\bffam=\sevenbf
     \def\tt{\fam\ttfam \twelvett \f@ntkey=7 }\textfont\ttfam=\twelvett
     \def\caps{\fam\cpfam \twelvecp \f@ntkey=8 }\textfont\cpfam=\twelvecp
     \setbox\strutbox=\hbox{\vrule height 12pt depth 5pt width\z@}
     \samef@nt}

 \def\twelvepoint{\relax
     \textfont0=\twelverm          \scriptfont0=\ninerm
     \scriptscriptfont0=\sixrm
      \def\rm{\fam0 \twelverm \f@ntkey=0 }\relax
     \textfont1=\twelvei           \scriptfont1=\ninei
     \scriptscriptfont1=\sixi
      \def\oldstyle{\fam1 \twelvei\f@ntkey=1 }\relax
     \textfont2=\twelvesy          \scriptfont2=\ninesy
     \scriptscriptfont2=\sixsy
     \textfont3=\twelveex          \scriptfont3=\elevenex
     \scriptscriptfont3=\tenex
     \def\it{\fam\itfam \twelveit \f@ntkey=4 }\textfont\itfam=\twelveit
     \def\sl{\fam\slfam \twelvesl \f@ntkey=5 }\textfont\slfam=\twelvesl
     \scriptfont\slfam=\ninesl
     \def\bf{\fam\bffam \twelvebf \f@ntkey=6 }\textfont\bffam=\twelvebf
     \scriptfont\bffam=\ninebf     \scriptscriptfont\bffam=\sixbf
     \def\tt{\fam\ttfam \twelvett \f@ntkey=7 }\textfont\ttfam=\twelvett
     \def\caps{\fam\cpfam \twelvecp \f@ntkey=8 }\textfont\cpfam=\twelvecp
     \setbox\strutbox=\hbox{\vrule height 10pt depth 4pt width\z@}
     \samef@nt}

 \def\tenpoint{\relax
     \textfont0=\tenrm          \scriptfont0=\sevenrm
     \scriptscriptfont0=\fiverm
     \def\rm{\fam0 \tenrm \f@ntkey=0 }\relax
     \textfont1=\teni           \scriptfont1=\seveni
     \scriptscriptfont1=\fivei
     \def\oldstyle{\fam1 \teni \f@ntkey=1 }\relax
     \textfont2=\tensy          \scriptfont2=\sevensy
     \scriptscriptfont2=\fivesy
     \textfont3=\tenex          \scriptfont3=\tenex
     \scriptscriptfont3=\tenex
     \def\it{\fam\itfam \tenit \f@ntkey=4 }\textfont\itfam=\tenit
     \def\sl{\fam\slfam \tensl \f@ntkey=5 }\textfont\slfam=\tensl
     \def\bf{\fam\bffam \tenbf \f@ntkey=6 }\textfont\bffam=\tenbf
     \scriptfont\bffam=\sevenbf     \scriptscriptfont\bffam=\fivebf
     \def\tt{\fam\ttfam \tentt \f@ntkey=7 }\textfont\ttfam=\tentt
     \def\caps{\fam\cpfam \tencp \f@ntkey=8 }\textfont\cpfam=\tencp
     \setbox\strutbox=\hbox{\vrule height 8.5pt depth 3.5pt width\z@}
     \samef@nt}



 \newtoks\date
 \def\monthname{\relax\ifcase\month 0/\or January\or February\or
    March\or April\or May\or June\or July\or August\or September\or
    October\or November\or December\else\number\month/\fi}
 \date={October 6, 1992}
 \def\today{\the\day\ \monthname\ \the\year}
 \def\ie{\hbox{\it i.e.}}

 \def\\{\relax\ifmmode\backslash\else$\backslash$\fi}
 \def\nextline{\unskip\nobreak\hskip\parfillskip\break}
 
 \def\journal#1&#2,#3(#4){\unskip \enskip {\sl #1}~{\bf #2}, {\rm #3 (19#4)}}

 \def\topspace{\hrule height 0pt depth 0pt \vskip}
\def\nullbox#1#2#3{\vbox to #1
     {\vss\vtop to#2
       {\vss\hbox to #3 {}}}}
\def\n@ll{\nullbox{5pt}{3pt}{2pt}}    
\def\UNRTXT#1{\vtop{\hbox{#1}\kern 1pt \hrule}}
\def\undertext#1{\ifvmode\ifinner \UNRTXT{#1}
                         \else    $\hbox{\UNRTXT{#1}}$
                         \fi
                 \else   \ifmmode \hbox{\UNRTXT{#1}}
                         \else    \UNRTXT{#1}
                         \fi
                 \fi }



\def\low#1{\kern-0.11em\lower0.4em\hbox{$\scriptstyle #1 $}\hskip-0.08em}
\def\up#1{\kern-0.45em\raise0.4em\hbox{$\scriptstyle #1 $}\hskip0.29em}


 \let\int=\intop         
 \def\prop{\mathrel{{\mathchoice{\pr@p\scriptstyle}{\pr@p\scriptstyle}{
                 \pr@p\scriptscriptstyle}{\pr@p\scriptscriptstyle} }}}
 \def\pr@p#1{\setbox0=\hbox{$\cal #1 \char'103$}
    \hbox{$\cal #1 \char'117$\kern-.4\wd0\box0}}
 \let\sec@nt=\sec
 \def\sec{\relax\ifmmode\let\n@xt=\sec@nt\else\let\n@xt\section\fi\n@xt}
 \def\@versim#1#2{\lower0.2ex\vbox{\baselineskip\z@skip\lineskip\z@skip
   \lineskiplimit\z@\ialign{$\m@th#1\hfil##\hfil$\crcr#2\crcr\sim\crcr}}}
\def\lsim{\mathrel{\mathpalette\@versim<}}
\def\gsim{\mathrel{\mathpalette\@versim>}}



\def\part#1#2{\partial^{#1}\kern-.1667em\hbox{$#2$}}
\def\der#1#2{d^{#1}\kern-.1067em\hbox{$#2$}}

\def\delt#1{{\delta\kern-.15em #1 }}



\def\rightbracearrow{\raise0.5em\hbox{$|$}\mkern-8.0mu\rightarrow}


\def\U#1{U\kern-.25em \left( #1 \right) }




 \normalbaselineskip = 20pt plus 0.2pt minus 0.1pt
 \normallineskip = 1.5pt plus 0.1pt minus 0.1pt
 \normallineskiplimit = 1.5pt
 \newskip\normaldisplayskip
    \normaldisplayskip = 20pt plus 5pt minus 10pt
 \newskip\normaldispshortskip
    \normaldispshortskip = 6pt plus 5pt
 \newskip\normalparskip
    \normalparskip = 6pt plus 2pt minus 1pt
 \newskip\skipregister
    \skipregister = 5pt plus 2pt minus 1.5pt
 \newif\ifsingl@    \newif\ifdoubl@
 \newif\ifb@@kspace   \b@@kspacefalse
 \newif\iftwelv@    \twelv@true
 \newif\ifp@genum


 \def\singlespace{\singl@true\doubl@false\b@@kspacefalse\spaces@t}
 \def\doublespace{\singl@false\doubl@true\b@@kspacefalse\spaces@t}
 \def\normalspace{\singl@false\doubl@false\b@@kspacefalse\spaces@t}
 \def\bookspace{\b@@kspacetrue\singl@false\doubl@false\spaces@t}
 \def\tablespace{\singlespace}
 \def\smashspace{\singl@false\doubl@false\b@@kspacefalse\subspaces@t3:8;}


 \def\Tenpoint{\tenpoint\twelv@false\spaces@t}
 \def\Twelvepoint{\twelvepoint\twelv@true\spaces@t}


 \def\spaces@t{\relax
       \iftwelv@ \ifsingl@\subspaces@t3:4;\else\subspaces@t1:1;\fi
        \else \ifsingl@\subspaces@t1:2;\else\subspaces@t4:5;\fi \fi
       \ifb@@kspace
          \baselineskip=14pt
       \fi
       \ifdoubl@ \multiply\baselineskip by 5
          \divide\baselineskip by 4 \fi }
 \def\subspaces@t#1:#2;{
       \baselineskip = \normalbaselineskip
       \multiply\baselineskip by #1 \divide\baselineskip by #2
       \lineskip = \normallineskip
       \multiply\lineskip by #1 \divide\lineskip by #2
       \lineskiplimit = \normallineskiplimit
       \multiply\lineskiplimit by #1 \divide\lineskiplimit by #2
       \parskip = \normalparskip
       \multiply\parskip by #1 \divide\parskip by #2
       \abovedisplayskip = \normaldisplayskip
       \multiply\abovedisplayskip by #1 \divide\abovedisplayskip by #2
       \belowdisplayskip = \abovedisplayskip
       \abovedisplayshortskip = \normaldispshortskip
       \multiply\abovedisplayshortskip by #1
         \divide\abovedisplayshortskip by #2
       \belowdisplayshortskip = \abovedisplayshortskip
       \advance\belowdisplayshortskip by \belowdisplayskip
       \divide\belowdisplayshortskip by 2
       \smallskipamount = \skipregister
       \multiply\smallskipamount by #1 \divide\smallskipamount by #2
       \medskipamount = \smallskipamount \multiply\medskipamount by 2
       \bigskipamount = \smallskipamount \multiply\bigskipamount by 4 }
 \def\normalbaselines{ \baselineskip=\normalbaselineskip
    \lineskip=\normallineskip \lineskiplimit=\normallineskip
    \iftwelv@\else \multiply\baselineskip by 4 \divide\baselineskip by 5
      \multiply\lineskiplimit by 4 \divide\lineskiplimit by 5
      \multiply\lineskip by 4 \divide\lineskip by 5 \fi }



 \newskip\tablelineskip           \tablelineskip=0.7in
 \newskip\figurelineskip          \figurelineskip=1.55in
 \newskip\tablelinelength     \tablelinelength=4.5in
 \newskip\tabledotvskip       \tabledotvskip=-0.359in
 \newskip\tabledothskip       \tabledothskip=-0.306in%
 \newskip\bookchapterskip     \bookchapterskip=1.0in
 \newcount\chapternumber    \chapternumber=0
 \newcount\appendixnumber   \appendixnumber=0
 \newcount\sectionnumber    \sectionnumber=0
 \newcount\subsectionnumber \subsectionnumber=0
 \newcount\equanumber       \equanumber=1
 \newcount\problemnumber    \problemnumber=0
 \newcount\figurecount      \figurecount=1
 \newcount\conpage          \conpage=0
 \let\chapterlabel=0
 \newtoks\constyle          \constyle={\Number}
 \newtoks\appendixstyle     \global\appendixstyle={\Alphabetic}
 \newtoks\chapterstyle      \chapterstyle={\Number}
 \newtoks\subsecstyle       \subsecstyle={\alphabetic}
 \newskip\chapterskip       \chapterskip=\bigskipamount
 \newskip\sectionskip       \sectionskip=\medskipamount
 \newskip\headskip          \headskip=8pt plus 3pt minus 3pt
 \newif\ifsp@cecheck
 \newif\iffirst@ppendix       \global\first@ppendixtrue
 \newdimen\chapterminspace    \chapterminspace=15pc
 \newdimen\sectionminspace    \sectionminspace=8pc
 \interlinepenalty=50
 \interfootnotelinepenalty=5000
 \predisplaypenalty=9000
 \postdisplaypenalty=500
 \newwrite\tableconwrite
 \newbox\tableconbox
 \newif\iftabl@conlist
 \newwrite\figwrite
 \newwrite\figurewrite
 \newif\iffigur@list
 \newwrite\eqnwrite
 \newif\if@qlist
 \newif\ifeqlo@d            \eqlo@dfalse
 \newwrite\tablewrite
 \newwrite\tableswrite
 \newif\iftabl@list
 \newwrite\appendixwrite
 \newif\if@ppendix
 \newcount\@ppcharnumber     \@ppcharnumber=64
 \newcount\referencecount \newbox\referencebox
 \newcount\tablecount
 \newif\ifindex      \indexfalse
 \newwrite\indexwrite
 \newbox\indexbox
 \newskip\indexskip
 \newif\ifmanualpageno   \manualpagenofalse
 \newdimen\letterhsize       \letterhsize=6.5in
 \newdimen\lettervsize       \lettervsize=8.5in
 \newdimen\labelhsize        \labelhsize=8.5in
 \newdimen\labelvsize        \labelvsize=11.0in
 \newdimen\letterhoffset     \letterhoffset=0.0in
 \newdimen\lettervoffset     \lettervoffset=0.250in


 \def\Number#1{\number #1}
 \def\makel@bel{\xdef\chapterlabel{%
     \the\chapterstyle{\the\chapternumber}.}}
 \def\sectionlabel{\number\sectionnumber }
 \def\subseclabel{{\the\subsecstyle{\the\subsectionnumber}. }}
 \def\alphabetic#1{\count255='140 \advance\count255 by #1\char\count255}
 \def\Alphabetic#1{\count255=64 \advance\count255 by #1\char\count255}
 \def\Roman#1{\uppercase\expandafter{\romannumeral #1}}


 \countdef\pagenumber=1  \pagenumber=1
 \def\advancepageno{\global\advance\pageno by 1
    \ifnum\pagenumber<0 \global\advance\pagenumber by -1
     \else\global\advance\pagenumber by 1 \fi \global\frontpagefalse }
\def\pagefolio#1{\ifnum#1<0 \romannumeral-#1
            \else \number#1 \fi }

\def\folio{\pagefolio{\pagenumber}}
 \def\pagecontents{
    \ifvoid\topins\else\unvbox\topins\vskip\skip\topins\fi
    \dimen@ = \dp255 \unvbox255
    \ifvoid\footins\else\vskip\skip\footins\footrule\unvbox\footins\fi
    \ifr@ggedbottom \kern-\dimen@ \vfil \fi }
 \def\makeheadline{\vbox to 0pt{ \hfuzz=30pt \skip@=\topskip
       \advance\skip@ by -12pt \advance\skip@ by -2\normalbaselineskip
       \vskip\skip@ \line{\vbox to 12pt{}\the\headline\hfill} \vss
       }\nointerlineskip}
 \def\makefootline{\baselineskip = 1.5\normalbaselineskip
                  \line{\the\footline}}
 \def\nopagenumbers{\p@genumfalse}
 \def\pagenumbers{\p@genumtrue}


 \def\tableconbreakfill{\hfill\break}


 \let\conbreak=\tableconbreak



 \def\splitprep{\global\newlinechar=`\^^J}
 \def\splitprepend{\global\newlinechar=-1}

 \def\consection#1{
    \iftabl@conlist
       \splitprep
       \immediate\write\tableconwrite{\string\immediate
                                      \string\spacecheck\sectionminspace
                                     }
       \immediate\write\tableconwrite{\string\vskip0.25in
                         \string\titlestyle{\string\bf\ #1 }%
                         \string\vskip0.0in
                         \string\nullbox{1pt}{1pt}{1pt}%
                                     }
       \splitprepend
    \fi}
 \def\unnumberedchapters{\let\makel@bel=\relax \let\chapterlabel=\relax
             \let\sectionlabel=\relax \let\subseclabel=\relax \equanumber=-1 }
 \def\titlestyle#1{\par\begingroup \interlinepenalty=9999
      \leftskip=0.02\hsize plus 0.23\hsize minus 0.02\hsize
      \rightskip=\leftskip \parfillskip=0pt
      \hyphenpenalty=9000 \exhyphenpenalty=9000
      \tolerance=9999 \pretolerance=9000
      \spaceskip=0.333em \xspaceskip=0.5em
      \iftwelv@\fourteenpoint\else\twelvepoint\fi
    \noindent #1\par\endgroup }
 \def\spacecheck#1{\p@gecheck{#1}%
    \ifsp@cecheck
       \else \vfill\eject
    \fi}
 \def\majorreset{
    \ifnum\figurecount<0\else\global\figurecount=1\fi
    \ifnum\equanumber<0 \else\global\equanumber=1\fi
    \sectionnumber=0 \subsectionnumber=0
    \tablecount=0  \problemnumber=0
    \bookheadline={}%
    \chapterheadline={}%
    }
 \def\chapterreset{\global\advance\chapternumber by 1
                   \majorreset
                   \makel@bel}
 \def\appendflag#1{
        \if@ppendix
        \else
          \global\@ppendixtrue
          \starttable{APPENDICES CALLED}{\appendixout}{\appendixwrite}{6}
        \fi
        \@ddconentry{\appendixwrite}{\noindent{\bf #1}}{1}
                   }



 \def\Textindent#1{\noindent\llap{#1\enspace}\ignorespaces}


 \newtoks\temphold
 \newtoks\ta
 \newtoks\tb
 \newtoks\captiontoks
 \newwrite\capwrite
 \newcount\runner  \runner=0
 \newcount\wordsnum \wordsnum=0
\newif\iflongc@p  \longc@pfalse
\def\longcap{\bgroup\obeyspaces\endlinechar=-1
             \global\longc@ptrue}
\def\endlongcap{\egroup\global\longc@pfalse}

 \def\confolio{\pagefolio{\conpage}}
 \def\p@gecheck#1{
        \dimen@=\pagegoal
        \advance\dimen@ by -\pagetotal
        \ifdim\dimen@<#1
           \global\sp@cecheckfalse
           \else\global\sp@cechecktrue
        \fi}
 \def\st@rtt@ble#1#2#3{
                 \message{listing #1:  type
                          \noexpand#2 for list}
                 \global\setbox#3=\vbox{\normalbaselines
                      \titlestyle{\seventeenrm\bf #1} \vskip\headskip}}
 \def\starttable#1#2#3#4{%
     \message{external listing of #1:  type %
              \noexpand#2 for list}%
     \ifcase#4%
           \immediate\openout#3=TABLE_OF_CONTENTS.TEX
       \or%
           \immediate\openout#3=FIG_CAP_AND_PAGE.TEX
       \or%
           \immediate\openout#3=FIGURE_CAPTIONS.TEX
       \or%
           \immediate\openout#3=TABLE_CAPTIONS.TEX
       \or%
           \immediate\openout#3=TAB_CAP_AND_PAGE.TEX
       \or%
           \immediate\openout#3=EQN_PAGE.TEX
       \or%
           \immediate\openout#3=APP_CALLED.TEX
       \or%
           \immediate\openout#3=REFERENC.TEX
      \else \immediate\message{You are in big trouble call a %
                                 TeXnician}%
     \fi%
     \ifIEEE
        \ifnum#4=7\immediate\write#3{\string\vbox {\string\normalbaselines%
              \string\bf\noindent References \string\vskip \string\headskip}}
        \fi%
     \else\immediate\write#3{\string\vbox {\string\normalbaselines%
               \string\titlestyle {\string\seventeenrm \string\bf \ #1}
               \string\vskip \string\headskip}}\fi%
 }%

\def\untoken#1#2{
     \ta=\expandafter{#1}\tb=\expandafter{#2}%
     \immediate\edef\temphold{\the\ta\the\tb}
     \global\captiontoks=\expandafter{\temphold}}%

\def\getc@pwrite#1{
                   \ifx#1\endcaption%
                      \let\next=\relax%
                      \immediate\write\capwrite{\the\captiontoks}
                      \global\captiontoks={}
                    \else%
                       \untoken{\the\captiontoks}{#1}
                       \ifx#1\space%
                             \advance\runner by 1
                             \advance\wordsnum by1 \let\next=\getc@pwrite
                       \fi%
                    \let\next=\getc@pwrite\fi%
                 \ifnum\runner=10
                       \immediate\message{.}
                       \immediate\write\capwrite{\the\captiontoks}
                       \global\captiontoks={}
                       \runner=0
                 \fi
                       \next}%

\def\c@pwrite#1#2{
               \let\capwrite=#1
               \global\runner=0
               \immediate\message{working on long caption. .}
               \wordsnum=0 \getc@pwrite#2\endcaption
               \immediate\message{The number of words= \number\wordsnum}}

\def\t@@bl@build@r#1#2{\relax
         \let\conbreak=\tableconbreakfill 
         \conpage=#2
         \t@mp=\tablelineskip
         \multiply\t@mp by #1
         \@@@@@=\tablelinelength \advance\@@@@@ by -\t@mp%
                                 \advance\@@@@@ by -1in%
             \@@@@@two=\tablelinelength%
             \advance\@@@@@two by -1in%
             \vglue\separationskip
         \vbox\bgroup
             \vbox\bgroup\parshape=3 0pt\@@@@@two%
                               \t@mp\@@@@@%
                               \t@mp\@@@@@%
                   \parfillskip=0pt%
                   \pretolerance=9000 \tolerance=9999 \hbadness=2000%
                   \hyphenpenalty=9000 \exhyphenpenalty=9000%
                   \interlinepenalty=9999%
                   \tablespace
                   \vskip\baselineskip\raggedright\noindent}
\def\endt@@bl@build@r{{\tabledotfill \hskip-0.07in $\,$}%
                  \egroup%
             \parshape=0%
            \vskip\tabledotvskip\hskip\@@@@@two\hskip\tabledothskip%
            { \tabledotfill\quad} {\confolio}%
         \egroup
         \vskip-0.10in}%

 \def\addconentry#1#2{%
     \setbox0=\vbox{\normalbaselines #2}%
     \relax%
     \global\setbox#1=\vbox{\unvbox #1 \vskip 4pt plus 2pt minus 1pt \box0 }}%

\def\@ddconentry#1#2#3{
     \conpage=\pagenumber%
     \p@gecheck{0pt}%
     \ifsp@cecheck \else
         \if\conpage<0
               \advance\conpage by -1
            \else
               \advance\conpage by 1
         \fi%
     \fi%
     \let\conbreak=\tableconbreakfill
     \splitprep
     \immediate\write#1
              {\string\t@@bl@build@r}%
     \immediate\write#1{{#3}%
                        {\the\conpage}%
                       }%
     \iflongc@p
         \immediate\c@pwrite{#1}{ #2 }%
       \else
         \immediate\write#1{{#2}}%
     \fi
     \immediate\write#1
              {\string\endt@@bl@build@r}%
     \splitprepend
}

\def\@ddfigure#1#2#3#4{
        \splitprep
        \captiontoks={}
        \immediate\write#1{{\string\vglue\string\separationskip}%
                          }%
        \immediate\write#1
              {\string\par \string\nullbox {1pt}{1pt}{1pt}%
              }%
        \immediate\write#1{\string\vbox}%
        \immediate\write#1{\string\bgroup \string\tablespace}
        \immediate\write#1{\string\parshape=2}
        \immediate\write#1{0pt\string\tablelinelength}
        \immediate\write#1{\string\figurelineskip\string\@@@@@}
        \immediate\write#1{\string\bgroup
                           }
       \iflongc@p
            \immediate\c@pwrite{#1}{#4~#2:\quad\ #3}
          \else
            \immediate\write#1{#4~#2:\quad\ #3}
       \fi
       \immediate\write#1{\string\egroup
                          \string\egroup
                         }%
       \immediate\write#1{\string\parshape=0%
                          }%
       \splitprepend
                     }%

 \def\captionsetup{
        \@@@@@=\tablelinelength \advance\@@@@@ by -\figurelineskip}
 \def\manualpageno{\manualpagenotrue}
 
 \def\dumplist#1{%
     \bookheadline={}%
     \chapterheadline={}%
     \unlock   
     \global\let\conbreak=\tableconbreakfill
     \ifmanualpageno
         \else
            \pagenumber=1
     \fi
     \ifb@@kstyle
        \global\multiply\pagenumber by -1
        \bookheadline={}
     \fi
     \vskip\chapterskip  
     \ifcase#1
            \chapterheadline={Contents}
            \bookheadline={Contents}    
            \input TABLE_OF_CONTENTS.TEX
       \or
            \chapterheadline={Contents}
            \bookheadline={Contents}         
            \input User_table_of_contents.tex
       \or
            \chapterheadline={Contents}
            \bookheadline={Contents}         
            \input User_table_of_contents.tex
            \input TABLE_OF_CONTENTS.TEX
       \or
           {\obeylines
            \input index.tex
           }
       \or
           {\obeylines
            \input User_index.tex
           }
       \or
           {\obeylines
            \input User_index.tex
            \input index.tex%
           }
       \or
           \captionsetup
           \input FIGURE_CAPTIONS.TEX
       \or
           \input FIG_CAP_AND_PAGE.TEX
       \or
           \captionsetup
           \input TABLE_CAPTIONS.TEX
       \or
           \input TAB_CAP_AND_PAGE.TEX
       \or
           \input EQN_PAGE.TEX
       \or
           \input APP_CALLED.TEX
       \or
           \input REFERENC.TEX
       \else \immediate\message{ Call a TeXnician, You are in BIG trouble}
     \fi
     \vfill\eject 
     \lock}
 \def\dumpbox#1{         
     \ifb@@kstyle
        \multiply\pagenumber by -1
        \bookheadline={}
     \fi
     \vskip\chapterskip  
     \unvbox#1           
     \vfill\eject}       


 \def\notabledots{\global\let\tabledotfill=\hfill}
 \def\tabledots{\global\let\tabledotfill=\dotfill}
 \def\tableconlist{\global\tabl@conlisttrue
                   \starttable{CONTENTS}{\conout}{\tableconwrite}
                              {0}}
 \def\tableconlistoff{\global\tabl@conlistfalse
                      \immediate\closeout\tableconwrite}

 \def\reflist{\global\referencelisttrue}

 \def\figurelist{\global\figur@listtrue
                 \starttable{FIGURE CAPTIONS AND PAGES}
                       {\figuresout}{\figurewrite}{1}
                 \starttable{FIGURE CAPTIONS}{\figout}{\figwrite}{2}
                }
 \def\figurelistoff{\global\figur@listfalse}

 \def\tablelist{\global\tabl@listtrue
                \starttable{TABLE CAPTIONS}{\tabout}{\tablewrite}{3}
                \starttable{TABLE CAPTIONS AND PAGES}
                           {\tablesout}{\tableswrite}{4}
               }
 \def\tablelistoff{\global\tabl@listfalse}


 \def\produceequations#1{
     \ifnum\equanumber<0
          \immediate\write\equ@tions{\string\xdef \string#1
                  {{\string\rm  (\number-\equanumber)}} }
       \else
          \immediate\write\equ@tions{\string\xdef \string#1
                  {{\string\rm  (\chapterlabel\number\equanumber)}} }
     \fi}


 \def\equ@tionlo@d{ \ifeqlo@d
                       \else \input BOOK_EQUATIONS
                             \eqlo@dtrue
                       \fi}


 \def\appendixout{\immediate\closeout\appendixwrite
                  \dumplist{11}
                 }
 \def\conout{\normalspace\immediate\closeout\tableconwrite
             \dumpbox{\tableconbox}
             \dumplist{0}}

 \def\refout{\immediate\closeout\referencewrite
             \dumplist{12}
   }
 \def\figout{
             \immediate\closeout\figwrite
             \dumplist{6}}
 \def\figuresout{
             \normalspace\immediate\closeout\figurewrite
             \dumplist{7}}
 
 \def\tabout{\immediate\closeout\tablewrite
             \dumplist{8}}
 \def\tablesout{\normalspace\immediate\closeout\tableswrite
          \dumplist{9}}
 
 \def\eqout{\immediate\closeout\eqnwrite
          \dumplist{10}
          \immediate\closeout\equ@tions}



\newdimen\referenceminspace  \referenceminspace=25pc
\newcount\referencecount     \referencecount=0
\newcount\titlereferencecount     \titlereferencecount=0
\newcount\lastrefsbegincount \lastrefsbegincount=0
\newdimen\refindent     \refindent=30pt
\newif\ifreferencelist       \global\referencelisttrue
\newif\ifreferenceopen       \global\referenceopenfalse
\newwrite\referencewrite
\newwrite\equ@tions
\newtoks\rw@toks


\def\NPrefmark#1{\attach{\scriptscriptstyle [ #1 ] }}
\def\PRrefmark#1{\attach#1}
\def\IEEErefmark#1{ [#1]}
\def\refmark#1{\relax%
     \ifPhysRev\PRrefmark{#1}%
     \else%
        \ifIEEE\IEEErefmark{#1}%
          \else%
            \NPrefmark{#1}%
          \fi%
     \fi}


\def\REF#1#2{\space@ver{}\refch@ck
   \global\advance\referencecount by 1 \xdef#1{\the\referencecount}%
   \r@fwrit@{#1}{#2}}%
\def\ref#1{\REF\?{#1}\refend}%
\def\Ref#1#2{\REF#1{#2}\refend}%
\def\refend{\global\lastrefsbegincount=\referencecount
            \refsend}
\def\REFS#1#2{\REF#1{#2}\global\lastrefsbegincount=\referencecount\relax}%
\def\REFSCON{\REF}%
\def\refsend{\refmark{\count255=\referencecount 
   \advance\count255 by-\lastrefsbegincount %
   \ifcase\count255 %
       \number\referencecount
   \or%
      \number\lastrefsbegincount,\number\referencecount
   \else%
      \number\lastrefsbegincount -\number\referencecount
  \fi}}%


\def\refch@ck{\ifreferencelist%
                  \ifreferenceopen%
                       \else \global\referenceopentrue%
                           \starttable{REFERENCES}{\refout}{\referencewrite}{7}
                  \fi%
              \fi%
}


\def\rw@begin#1\splitout{\rw@toks={#1}\relax%
   \immediate\write\referencewrite{\the\rw@toks}\futurelet\n@xt\rw@next}%
\def\rw@next{\ifx\n@xt\rw@end \let\n@xt=\relax%
      \else \let\n@xt=\rw@begin \fi \n@xt}%
\let\rw@end=\relax%
\let\splitout=\relax%
\def\r@fwrit@#1#2{%
   \splitprep%
   \immediate\write\referencewrite{\ifIEEE\noexpand\refitem{\noindent[#1]}
                                   \else\noexpand\refitem{#1.}\fi}%
   \rw@begin #2\splitout\rw@end \@sf%
   \splitprepend%
                 }%


\def\refitem#1{\par \hangafter=0 \hangindent=\refindent \Textindent{#1}}%



%
%


\def\TITLEREF#1#2{\space@ver{}\refch@ck%
   \global\advance\titlereferencecount by 1%
   \xdef#1{\alphabetic{\the\titlereferencecount}}%
   \r@fwrit@{#1}{#2}}%
%



 \def\checkreferror{\ifinner\errmessage{This is a wrong place to DEFINE
     a reference or a figure! You may have your references / figure
     captions screwed up. }\fi}


 \def\FIG#1#2{\checkreferror
     \ifnum\figurecount<0
          \global\xdef#1{\number-\figurecount}%
          \global\advance\figurecount by -1
       \else
          \global\xdef#1{\chapterlabel\number\figurecount}%
          \global\advance\figurecount by 1
     \fi
     \iffigur@list
        \@@@@@=\hsize \advance\@@@@@ by -1.219in
        \@ddconentry{\figurewrite}%
            {\noindent
                \bgroup \bf \expandafter{\csname\string#1.\endcsname} \egroup
                \quad{#1}\quad\ #2
            }
            {1}%
        \@ddfigure{\figwrite}{#1}{#2}{Figure}%
     \fi     }%



 \def\nexttable{\checkreferror\global\advance\tablecount by 1}


 \def\TABLE#1#2{\nexttable\xdef#1{\chapterlabel\the\tablecount}%
     \iftabl@list
        \@ddfigure{\tablewrite}{#1}{#2}{Table}%
        \@ddconentry{\tableswrite}%
               {\noindent
                   \bgroup \bf \expandafter{\csname\string#1.\endcsname}
\egroup
                   \quad{#1}\quad\ #2}
               {1}%
     \fi}%



 \def\eqname#1{\relax
     \if@qlist
       \produceequations{#1}
     \fi
     \ifnum\equanumber<0
           \global\xdef#1{{\rm(\number-\equanumber)}}
           \global\advance\equanumber by -1
     \else
           \global\xdef#1{{\rm(\chapterlabel \number\equanumber)}}
           \global\advance\equanumber by 1
     \fi
     \if@qlist
       \@ddconentry{\eqnwrite}
           {\noindent
             \bgroup \bf \expandafter{\csname\string#1\endcsname.} \egroup
             \quad{#1}\quad}
           {1}
     \fi      }



 \def\eqinsert#1{\noalign{\dimen@=\prevdepth \nointerlineskip
    \setbox0=\hbox to\displaywidth{\hfil #1}
    \vbox to 0pt{\vss\hbox{$\!\box0\!$}\kern-0.5\baselineskip}
    \prevdepth=\dimen@}}



 \def\globaleqnumbers{\relax\if\equanumber<0\else\global\equanumber=-1\fi}


\def\globalfigurenumbers{\relax\if\figurecount<0\else\global\figurecount=-1\fi}



 \newskip\lettertopfil      \lettertopfil = 0pt plus 1.5in minus 0pt
 \newskip\spskip            \setbox0\hbox{\ } \spskip=-1\wd0
 \newskip\signatureskip     \signatureskip=40pt
 \newskip\letterbottomfil   \letterbottomfil = 0pt plus 2.3in minus 0pt
 \newskip\frontpageskip     \frontpageskip=1\medskipamount plus .5fil
 \newskip\headboxwidth      \headboxwidth= 0.0pt
 \newskip\letternameskip    \letternameskip=0pt
 \newskip\letterpush        \letterpush=0pt
 \newbox\headbox
 \newbox\headboxbox
 \newbox\physbox
 \newbox\letterb@x
 \newif\ifse@l              \se@ltrue
 \newif\iffrontpage
 \newif\ifletterstyle
 \newif\ifhe@dboxset        \he@dboxsetfalse
 \newtoks\memoheadline
 \newtoks\letterheadline
 \newtoks\letterfrontheadline
 \newtoks\lettermainheadline
 \newtoks\letterfootline
 \newdimen\holder
 \newdimen\headboxwidth
 \newtoks\myletterheadline  \myletterheadline={define \\myletterheadline}
 \newtoks\mylettername    \mylettername={{\twentyfourbf define}\
\\mylettername}
 \newtoks\phonenumber       \phonenumber={\rm (313) 764-4437}
 \newtoks\faxnumber	    \faxnumber={(313) 763-9694}
 \newtoks\telexnumber       \telexnumber={\tenfib 4320815 UOFM UI}


 \def\FIRSTP@GE{\ifvoid255\else\vfill\penalty-2000\fi
                \global\frontpagetrue}
 \def\FRONTPAGE{\ifvoid255\else\vfill\penalty-2000\fi
       \masterreset\global\frontpagetrue
       \global\lastp@g@no=-1 \global\footsymbolcount=0}
 
 \letterfootline={\hfil}
 \lettermainheadline={\hbox to \hsize{
        \rm\ifp@genum page \ \folio\fi
        \hfill\today}}
 \letterheadline{\hfuzz=60pt\iffrontpage\the\letterfrontheadline
      \else\the\lettermainheadline\fi}
 \def\addressee#1{\vskip-0.188in \singlespace
    \ialign to\hsize{\strut ##\hfill\tabskip 0pt plus \hsize \cr #1\hfill\crcr}
    \medskip\noindent\hskip\spskip}
 \def\letterstyle{\global\letterstyletrue%
                  \global\paperstylefalse%
                  \global\b@@kstylefalse%
                  \global\frontpagetrue%
                  \global\singlespace\global\lettersize}
 \def\lettersize{\hsize=\letterhsize \vsize=\lettervsize
                 \hoffset=\letterhoffset \voffset=\lettervoffset
    \headboxwidth=\hsize  \advance\headboxwidth by -1.05in
    \skip\footins=\smallskipamount \multiply\skip\footins by 3}
 \def\telex#1{\telexnumber={\tenfib #1}}
 \def\noseal{\se@lfalse}


 \def\myaddress#1{%
         \singlespace
         \global\he@dboxsettrue
         {\hsize=\headboxwidth \setbox0=\vbox{\ialign{##\hfill\cr #1 }}
         \global\setbox\headboxbox=\vbox {\hfuzz=50pt
                  \vfill
                  \line{\hfil\box0\hfil}}}}
 \def\MICH@t#1{\hfuzz=50pt\global\setbox\physbox=\hbox{\it#1}\hfuzz=1pt}
 \def\l@tt@rcheck{\ifhe@dboxset\else
                     \immediate\message{Setting default letterhead.}
                     \immediate\message{For more information consult the PHYZZM
                                       documents.}
                     \UM
                  \fi}
 \def\l@tt@rs#1{\l@tt@rcheck
              \vfill\supereject 
              \global\letterstyle
              \global\letterfrontheadline={}
              \global\setbox\headbox=\vbox{{
                  \hbox to\headboxwidth{\hepbig\hfil #1\hfil}
                  \vskip 1pc
                  \unvcopy\headboxbox}}}
 \def\HEAD#1{\singlespace  
    \global\letternameskip=22pt plus 0pt minus 0pt
    \global\memoheadline={UM\ #1\ Memorandum}
    \mylettername={\hep #1}
       \myaddress{
   \hep\hfill\hbox{The Harrison M. Randall Laboratory of Physics}\hfill \cr
   \hep\hfill\hbox{500 East University, Ann Arbor, Michigan 48109-1120}\hfill
\cr
                                                 \cr}}


 \def\theory{\phonenumber={\rm (313) 763-9698}
        \HEAD{Theoretical Physics}
        \PUBNUM{TH} \MICH@t{Theoretical Physics}}
 \def\UM{\phonenumber={\rm (313) 764-4437}
        \HEAD{Department of Physics}
        \PUBNUM{PHYSICS} \MICH@t{Department of Physics}}
 

 \def\lettertext{\par\unvcopy\letterb@x\par}
 \def\multiletter{\setbox\letterb@x=\vbox\bgroup
       \everypar{\vrule height 1\baselineskip depth 0pt width 0pt }
       \singlespace \topskip=\baselineskip }
 \def\letterend{\par\egroup}

 \def\letters{\l@tt@rs{The University of Michigan}}
 \def\letter{\wlog{\string\letter}
        \vfill\supereject
        \normalspace       
        \FRONTPAGE
        \nullbox{1pt}{1pt}{1pt}\vskip-0.8750in
        \setbox0 = \vbox{\hfuzz=50pt   
           {\hsize=\headboxwidth
	   \unvcopy\headbox\hfill
           \singlespace
           \vskip-0.1in
	   \dimen1=2.25truein
	   \setbox1=\hbox{\hep(313) 763-4929}     \advance\dimen1 by \wd1
	   \setbox1=\hbox{\hep\the\phonenumber}   \advance\dimen1 by -1\wd1
           \centerline{\the\mylettername\hskip\dimen1{\hep\the\phonenumber}}}}
%
%
 \ifse@l
   \dimen0=0.8in
   \setbox1=\hbox to \dimen0{%
    \dimen0=1.0in
    \vbox to \dimen0{
      \vss
        \UGbody{tex$inputs:umseal.ps}{1.0}{8.31}{1}                 
     }%
     \hss
    }
 \fi
        \hfuzz=5pt
        \ifse@l
          \hfuzz=15pt\line{\hbox to0pt{}\hskip 0.3in\box1\box0 \hfill}
	\else
	  \line{\hfill\box0\hfill}
	\fi
        \vskip0.35in
        \vskip\letterpush
        \rightline{\today}
        \addressee}

 \def\myletter{\l@tt@rs{\the\myletterheadline}\letter}

 \def\signed#1{\par \penalty 9000 \bigskip \dt@pfalse
   \everycr={\noalign{\ifdt@p\vskip\signatureskip\global\dt@pfalse\fi}}
   \setbox0=\vbox{\singlespace \halign{\tabskip 0pt \strut ##\hfill\cr
    \noalign{\global\dt@ptrue}#1\hfill\crcr}}
   \line{\hskip 0.5\hsize minus 0.5\hsize \box0\hfill} \medskip }
 \def\copies#1{\singlespace\hfill\break
   \line{\nullbox{0pt}{0pt}{\hsize}}
   \setbox0 = \vbox {
     \noindent{\tenrm cc:}\hfill 
     \vskip-0.2115in\hskip0.000in\vbox{\advance\hsize by-\parindent
       \ialign to\hsize{\strut ##\hfill\hfill
                 \tabskip 0pt plus \hsize \cr #1\crcr}}
       \hbox spread\hsize{}\hfill\vfill}
   \line{\box0\hfill}}
 \def\endletter{\nullbox{0pt}{0pt}{\hsize}
       \ifnum\pagenumber=1
                \vskip\letterbottomfil\vfill\supereject
          \else
                \vfill\supereject
       \fi
       \wlog{ENDLETTER}}



 \newif\ifp@bblock          \p@bblocktrue
 \newif\ifpaperstyle
 \newif\ifb@@kstyle
 \newtoks\paperheadline
 \newtoks\bookheadline
 \newtoks\chapterheadline
 \newtoks\paperfootline
 \newtoks\bookfootline
 \newtoks\Pubnum            \Pubnum={$\caps UM - PHY - PUB -
                                        \the\year - \the\pubnum $}
 \newtoks\pubnum            \pubnum={00}
 \newtoks\pubtype           \pubtype={\tensl Preliminary Version}
 \newcount\yeartest
 \newcount\yearcount

 \def\sequentialfootnotes{\global\seqf@@tstrue}
 \def\PHYSREV{\paperstyle\PhysRevtrue\PH@SR@V
     \let\refmark=\attach}
 \def\PH@SR@V{\doubl@true \baselineskip=24.1pt plus 0.2pt minus 0.1pt
              \parskip= 3pt plus 2pt minus 1pt }
 \def\IEEE{\paperstyle\IEEEtrue\I@EE\doublespace\rm\let\refmark=\IEEErefmark%
             \let\unnumberedchapters=\relax}
 \def\I@EE{\baselineskip=24.1pt plus 0.2pt minus 0.1pt
              \parskip= 3pt plus 2pt minus 1pt }


 \def\bookstyle{\b@@kstyletrue%
                \letterstylefalse%
                \paperstylefalse%
                \equ@tionlo@d
                \Tenpoint
                \frenchspacing
                \parskip=0pt
                \bookspace\booksize}
 \def\paperstyle{\paperstyletrue%
                 \b@@kstylefalse%
                 \letterstylefalse%
                 \normalspace}

\def\booksize{\hsize=29pc\vsize=45pc\hoffset=0.85in\voffset=0.475in\hfuzz=2.5pc
               \itemsize=\parindent}


 \paperfootline={\hss\iffrontpage
                         \else \ifp@genum
                                  \hbox to \hsize{\tenrm\hfill\folio\hfill}\hss
                               \fi
                      \fi}
 \paperheadline={\hfil}
 \bookfootline={\hss\iffrontpage
                       \ifp@genum
                              \hbox to \hsize{\tenrm\bf\hfill\folio\hfill}\hss
                       \fi
                     \else
                           \hbox to \hsize{\hfill\hfill}\hss
                     \fi}


 \def\titlepage{
    \yeartest=\year \advance\yeartest by -1900
    \ifnum\yeartest>\yearcount
          \global\PUBNUM{UM}
    \fi
    \FRONTPAGE\paperstyle\ifPhysRev\PH@SR@V\fi
    \ifp@bblock\p@bblock\fi}

 \def\nopubblock{\p@bblockfalse}
 
 \def\p@bblock{\begingroup \tabskip=\hsize minus \hsize
    \baselineskip=1.5\ht\strutbox \topspace-2\baselineskip
    \halign to\hsize{\strut ##\hfil\tabskip=0pt\crcr
    \the\Pubnum\cr \the\date\cr \the\pubtype\cr}\endgroup}


 \def\PUBNUM#1{
     \yearcount=\year
     \advance\yearcount by -1900
     \Pubnum={$\caps UM- #1 - \the\yearcount - \the\pubnum $}}
 \def\title#1{\vskip\frontpageskip \titlestyle{#1} \vskip\headskip }
 \def\author#1{\vskip\frontpageskip\titlestyle{\twelvecp #1}\nobreak}

 \def\address#1{\par\noindent\titlestyle{\twelvepoint\it #1}}

 \def\andaddress{\par\kern 5pt \centerline{\sl and} \address}
 \def\abstract{\vskip\frontpageskip\centerline{\fourteenrm ABSTRACT}
               \vskip\headskip }




 \newtoks\foottokens       \foottokens={\Tenpoint\singlespace}
 \newdimen\footindent      \footindent=24pt
 \newcount\lastp@g@no	   \lastp@g@no=-1
 \newcount\footsymbolcount \footsymbolcount=0
 \newif\ifPhysRev
 \newif\ifIEEE
 \newif\ifseqf@@ts         \global\seqf@@tsfalse



 \def\footrule{\dimen@=\prevdepth\nointerlineskip
    \vbox to 0pt{\vskip -0.25\baselineskip \hrule width 0.35\hsize \vss}
    \prevdepth=\dimen@ }
 \def\vfootnote#1{\insert\footins\bgroup  \the\foottokens
    \interlinepenalty=\interfootnotelinepenalty \floatingpenalty=20000
    \splittopskip=\ht\strutbox \boxmaxdepth=\dp\strutbox
    \leftskip=\footindent \rightskip=\z@skip
    \parindent=0.5\footindent \parfillskip=0pt plus 1fil
    \spaceskip=\z@skip \xspaceskip=\z@skip
    \Textindent{$ #1 $}\footstrut\futurelet\next\fo@t}


 \def\footnote#1{\attach{#1}\vfootnote{#1}}


 \let\footsymbol=\star
 \def\footsymbolgen
 {  \relax
   \ifseqf@@ts \seqf@@tgen
   \else
      \ifPhysRev
         \iffrontpage \NPsymbolgen
         \else \PRsymbolgen
         \fi
      \else \NPsymbolgen
      \fi
   \fi
   \footsymbol
 }
 \def\seqf@@tgen{\ifnum\footsymbolcount>0 \global\footsymbolcount=0\fi
       \global\advance\footsymbolcount by -1
       \xdef\footsymbol{\number-\footsymbolcount} }
 \def\NPsymbolgen
 {  \ifnum\footsymbolcount<0 \global\footsymbolcount=0\fi
   {  \iffrontpage \relax
      \else
         \ifnum \lastp@g@no = \pageno
            \relax
         \else
            \global\lastp@g@no = \pageno
            \global\footsymbolcount=0
         \fi
      \fi
   }
   \ifcase\footsymbolcount
      \fd@f\star \or \fd@f\dagger \or \fd@f\ast \or
      \fd@f\ddagger \or \fd@f\natural \or \fd@f\diamond \or
      \fd@f\bullet \or \fd@f\nabla
   \fi
   \global\advance\footsymbolcount by 1
   \ifnum\footsymbolcount>6 \global\footsymbolcount=0\fi
 }
 \def\fd@f#1{\xdef\footsymbol{#1}}
 \def\PRsymbolgen{\ifnum\footsymbolcount>0 \global\footsymbolcount=0\fi
       \global\advance\footsymbolcount by -1
       \xdef\footsymbol{\sharp\number-\footsymbolcount} }
 \def\space@ver#1{\let\@sf=\empty \ifmmode #1\else \ifhmode
    \edef\@sf{\spacefactor=\the\spacefactor}\unskip${}#1$\relax\fi\fi}
 \def\attach#1{\space@ver{\strut^{\mkern 2mu #1} }\@sf\ }


 \footline={\ifletterstyle \the\letterfootline
               \else \ifpaperstyle \the\paperfootline
                        \else \the\bookfootline
                      \fi
            \fi}
 \headline={\let\conbreak=\tableconbreakspace%
            \ifletterstyle \the\letterheadline
               \else \ifpaperstyle \the\paperheadline
                        \else \iffrontpage {}
                                  \else
                                     \setbox0=\hbox{\ {\tenrm\bf \folio}}
                                     \advance\hsize by \wd0
                                     \ifodd\pagenumber
                                          \hbox to \hsize{%
                                              \the\bookheadline\hfill\hfill
                                              \box0
                                                         }%
                                        \else
                                          \hskip-\wd0
                                          \hbox to \hsize{%
                                              \box0\hfill\hfill
                                              \the\chapterheadline
                                              }%
                                     \fi
                              \fi
                     \fi
             \fi}



 \def\masterreset{\global\pagenumber=1 \global\chapternumber=0
    \global\appendixnumber=0
    \global\equanumber=1 \global\sectionnumber=0 \global\subsectionnumber=0
    \global\referencecount=0 \global\figurecount=1 \global\tablecount=0
    \global\problemnumber=0
    \global\@ppendixfalse
    \ifIEEE\global\setbox\referencebox=\vbox{\normalbaselines
          \noindent{\bf References}\vskip\headskip}%
     \else\global\setbox\referencebox=\vbox{\normalbaselines
          \centerline{\fourteenrm REFERENCES}\vskip\headskip}\fi
   }




 \normalspace                      
 \masterreset
 \pagenumbers                      
 \Twelvepoint                      
 \figurelistoff                    
 \tableconlistoff                  
 \tablelistoff                     
 \reflist                          
 \telex{4320815 UOFM UI}           
 \headboxwidth=6.5in
 \advance\headboxwidth by -1.05in  
 \UM                               
 \paperstyle                       
 \tabledots                        
 \manualpageno                     
 \he@dboxsetfalse                  
 \lock
 \hfuzz=1pt
 \vfuzz=0.2pt

\message{PHYZZM version 1.9}

\def\un{\undertext}

\overfullrule=0pt
\pubnum{20}
\pubtype{}
\theory
\titlepage
\title{COSMIC RAYS, GAMMA RAYS AND NEUTRINOS FROM POINT
SOURCES}
\author{\caps Yukio Tomozawa}
\address{Randall Lab of Physics\nextline University of
Michigan\nextline
Ann Arbor, MI  48109-1120}
\abstract
The suggestion has been made that the energy spectrum from point
sources such as AGN (Active Galactic Nuclei) and GBHC (Galactic Black
Hole Candidates) is universal, irrespective of the nature of the
emitted particles.  A comparison of the energy spectrum for cosmic
rays at the source and $\gamma$-rays from quasars obtained recently by
CGRO (Compton Gamma Ray Observatory) indicates that the prediction is
in agreement with the data in the average sense.  This suggests that
neutrinos from point sources should have a spectral index identical to
that of $\gamma$-rays for an individual point source.  This prediction
is also consistent with the recent observation of neutrinos by
Kamiokande and IMB in which the ratio of $\nu_\mu/\nu_e$ is close to
1, instead of 2 as expected from atmospheric neutrinos.  For a
further test of the model, analysis of the time variation of
$\gamma$-ray spectra from quasars is suggested.
\smallskip
\centerline{\fourteenrm\bf I.\ \ Introduction}

The observation of high energy $\gamma$-rays from quasars (up to a few
GeV by EGRET of CGRO\REFS\one{R.C. Hartman et al., Ap. J.
\underbar{385} L1 (1992).}\REFSCON\two{E.J. Schneid et al., Astron.
Astrophys., to appear.}\refsend and up to a few TeV by the Whipple
Observatory\REF\three{M. Schubnell, Proc. Seventh International
Symposium on Very High Energy Cosmic Ray Interactions; Nature in
press.}\refend clearly calls for a drastically new approach for
understanding the $\gamma$-ray emission mechanism from point sources.
This is because electron infall onto a compact object generates X-rays
but not high energy $\gamma$-rays: (for such a process, (the
gravitational energy)/(the rest energy) $\lsim$ 1).  The purpose of
this article is to present a novel mechanism of $\gamma$-ray emission
from black holes and to compare its prediction with the observed data.
In fact, the model\REF\four{Y. Tomozawa, Proceedings of the INS
International Symposium on Composite Models of Quarks and Leptons (ed.
H. Terazawa and M. Yasue, 1985), p. 386.}\REFSCON\five{Y. Tomozawa, in
Quantum Field Theory (ed. F. Mancini, Elsevier Pub., 1986), p.
241.}\REFSCON\six{Y. Tomozawa, Lectures at the Second Workshop on
Fundamental Physics, Univ. of Puerto Rico, Humacao (ed. E. Estaban,
1986), p. 144.}\refmark{4-6} was proposed in 1985, much before the CGRO
operation, and the recent data is quite consistent with the
prediction.

The salient features of the theory can be summarized in the following
way.

\item{1.} The consideration of quantum effects on the Einstein
equation suggests that the gravitational potential is repulsive at
short distances. The rotation of black holes represented by the Kerr
metric has the similar feature, \ie\ the angular momentum plays the
role of a repulsive force, a phenomenon similar to the angular
momentum barrier in quantum mechanics.

\item{2.} Applying this result to gravitational collapse, one
encounters a novel phenomenon called black hole pulsation.  An
analysis of the behavior of black hole motion enables us to conclude
that the pulsation is observable.

\item{3.} The spectrum of particles emitted during the pulsation
is decided by the rate of expansion of the system.  This leads to the
prediction of a universal energy spectrum for different particles from
an individual pulsating compact object.  (There should be a variation
for the spectrum since an individual object can have a different
expansion rate depending on the environment.)

\item{4.} This prediction of universality can be compared with the data
of cosmic rays, $\gamma$-rays and neutrinos.  It will be shown that
the data is in reasonable agreement with the prediction and further
tests will be suggested.

Section II is devoted to a discussion of cosmic rays generated by
pulsating black holes, and $\gamma$-rays from point sources observed
by CGRO compared with the cosmic ray energy spectrum at the sources in
Section III.  Section IV presents a discussion on the neutrino
observation by the Kamiokande and IMB detectors, which may provide
evidence for the proposed model and Section V suggests the analysis of
time variation of the energy spectrum as a further test of the theory.
\smallskip
\centerline{\fourteenrm\bf II.\ \ Cosmic Rays}

Since the discovery of cosmic rays early this century, an impressive
amount of experimental data has been accumulated.  Yet, the origin of
cosmic rays defies the understanding of physicists.  Several important
questions are: What is the fraction of galactic and extragalactic
components of primary cosmic rays?  How can one understand the power
law energy spectrum $(\sim E^{-2.5} - E^{-3}$ for the energy range
$10^9\ \hbox{eV}\ < E < 10^{20}\ \hbox{eV}$)?  How do they attain such
high energies?  Despite various attempts in the past to answer these
questions, we are still left in the dark.\REF\longair{M.S. Longair,
{\it High Energy Astrophysics} (Cambridge University Press, 1981); S.
Hayakawa, {\it Cosmic Ray Physics} (Wiley-Interscience, NY,
1969).}\refend

It is known\REF\bell{A.R. Bell, M.N.R.S., \un{182}, 147, 443 (1978);
P.O. Lagage and C.J. Casarsky, Astr. Ap. \un{118}, 223 (1983); R.D.
Blandford and J.P. Ostriker, Ap. J. {\un{221,}} L29 (1978).}\refend
that shock wave acceleration in a supernova explosion does not explain
the high energy component of cosmic rays (above the so called knee
energy $10^{16}$ eV).  It was then suggested that strong magnetic
fields around pulsars may be responsible for the acceleration of high
energy cosmic rays\REF\gunn{J.E. Gunn and J.P. Ostriker, Phys. Rev.
Lett. 22, 728 (1969).}\refend above the knee energy.  In this case,
however, drastically different mechanisms are responsible for cosmic
ray acceleration below and above the knee energy.  This makes it
difficult to have the continuous spectrum observed in the experimental
data.  An alternative scheme is to invoke shock wave acceleration in
the galactic wind,\REF\jokipii{J.P. Jokipii and G.E. Morfill, Ap. J.
\un{290}, L1 (1985).}\refend the existence of which is yet to be
established by observation.

The application of BHP (black hole pulsation) leads naturally to the
emission of particles.  Since the temperature decreases with expansion
of quantum mechanical black holes or rotating black holes, the energy
spectrum of the emitted particles can be computed, provided the
expansion rate is known.

The number of particles of type $x$ emitted with energy $E$ is given
by
$$f_x (E) = {(2s+1)\over 2\pi^2} \int \eta_x (E/kT)\ {E^2 4\pi R^2 dt\over
e^{E/kT-\mu/kT} \pm 1},\eqno(1)$$
where $R$ is the radius of the system, $\eta_x (E/kT)dt$ is the fraction of
particles $x$ emitted in time interval $dt$ and $\mu$ and $s$ are the
chemical potential and the spin for particles of type
$x.$  The $+(-)$ sign in the denominator is for fermions (bosons).
Assume the relationship $R=a/kT$ and an expanding rate
$$t\ =\ bR^\alpha ,\eqno(2)$$
where $a$ and $b$ are constants.  The function $\eta_x
(E/kT)$ is unknown, but is assumed to scale as a function of $E/kT.$
The chemical potential for fermions is obtained by the condition
$${N\over V} = {(2s+1)\over 2\pi^2} (kT)^3\int^\infty_0\ {x^2dx\over
e^{x-\mu/kT} +1}\eqno(3)$$
or equivalently
$$\mu_0 \equiv \mu/kT = g\left({N\over V(kT)^3}\ {2\pi^2\over
(2s+1)}\right),\eqno(4)$$
where $N$ is the total number of particles in the system and $V$ is
the volume given by
$$V= {4\pi\over 3} R^3 = {4\pi\over 3}\ {a^3\over (kT)^3}.\eqno(5)$$
(At high temperature $g(x) = \ell n(x/2)$.)  Obviously, $\mu_0$ is
independent of temperature since $VT^3$ is constant during the course
of the expansion.  $(\mu_0 = 0$ for photons.)
Using Eqs. (1)-(4), we obtain
$$f_x(e) = {A_{x,\alpha}\over E^{\alpha}},\eqno(6)$$
where
$$A_{x,\alpha} = {2(2s+1)\alpha
b(a)^{2+\alpha}\over\pi}\>\int^\infty_0\> {\eta_x (s)s^{\alpha+1}
ds\over e^{s-\mu_0} \pm 1}\eqno(7)$$
is a constant.

Some discussion is in order.  First, how can one explain the observed
nuclei in cosmic rays ($\sim$10\% of primary cosmic rays which are
mostly protons below the knee energy).  One can invoke shock wave
acceleration in supernova explosions for 10-20 \% of cosmic rays.  Or
nuclei can be emitted from BHP, since the density is extremely high so
that the Fermi temperature is also extremely high.  As a result, the
situation can be like a low temperature state despite the temperature
$T$ is high.  Thus, due to the Boson condensation, even nuclei can be
emitted in this system.\REF\majumder{A. Majumder and Y. Tomozawa,
Prog. Theoret. Phys. \un{82}, 555 (1989).}\refend

Secondly, the cosmic ray energy spectrum observed at the earth may not
be the same as that at the sources.  Using leakage, spallation and
information on the chemical abundance of cosmic rays, the Chicago
group has derived the power index $\lambda$ of the cosmic ray energy
spectrum $E^{-\lambda}$ at the sources.  The most elaborate
analysis\Ref\swordy{S.P. Swordy et al., Ap. J. \underbar{349} (1990)
625; D. M\"uller et al., Ap. J.
\underbar{374} (1991) 356; P. Meyer et al., ICRC (1991)
OG6.1.11.}gives
$$\lambda_{\rm source} = 2.2 \pm 0.1.\eqno(8)$$
This index should be compared with the index for $\gamma$-ray energy
spectrum.

Finally, in our model the power law spectrum of cosmic rays is a
reflection of the power law expansion rate.  The knee energy is caused
by the difference of the expansion rate in the nonrelativistic and
relativistic regime.  Then, this would require the existence of the
energy scale of $\sim$ several hundred TeV which differentiates both
regimes.  However, the modification of the low energy spectrum with
Eq. (8) brings the knee energy at a much lower energy scale, around 1
TeV (instead of several hundred TeV).
\smallskip
\centerline{\fourteenrm \bf III.\ \ Gamma-Rays}

It is clear that the energy spectrum derived in the last section
applies for any particles emitted from BHP.  The important prediction
is, then

\noindent [Proposition]  Any particles emitted from point sources
such as GBHC or AGN should have identical energy spectrums on the
average, the universal spectrum, when they left the sources.

The most recent data by the EGRET detector gives the energy spectrum
for $\gamma$-rays from quasar 3C279 as $E^{-\lambda}$,
where\refmark{\one}
$$\lambda = 2.02 \pm 0,07,\eqno(9)$$
for the energy range 30 MeV to 5 GeV.  The same group
observed\refmark{\two} three bursts from the location at R.A. = 88.6
degrees and Dec. = 38.6 degrees. They have power index
$$\lambda = 2.13 \pm 0.08,\eqno(10a)$$
and
$$\lambda = 2.24 \pm 0.03.\eqno(10b)$$
The spectrum of the third burst is more complicated and it is
suggested as a composite of
$$\lambda = 2.22,\eqno(10c)$$
and
$$\lambda = 4.0 \pm 0.8\ \hbox{at the low energy end.}\eqno(11)$$
The proximity of the values of Eqs. (8), (9) and (10) seems to
support the concept of universality proposed by the author.  It should
be emphasize, however, that universality of the energy spectrum is
valid only in the sense of the average so that a certain amount of
fluctuation is inevitable.  As a matter of fact, the spectrum of
cosmic rays itself should be the outcome of an average of the spectrum
from many sources which power index has a fluctuation.
\smallskip
\centerline{\fourteenrm\bf IV.\ \ Muon Neutrino Deficiency and Cosmic
Neutrinos}

The concept of a universal energy spectrum from point sources can be
applied to other particle emissions.  An example is neutrinos.
Neutrinos are emitted from pulsating black holes with an intensity
comparable to that of $\gamma$-rays.  With the difference of
statistics and helicity, the neutrino spectrum is 3/8 of the
$\gamma$-ray spectrum, but they have the same spectral index (with the
same variation for individual sources, of course).  Moreover, the
intensity is the same for all species $\nu_e, \bar\nu_e, \nu_\mu,
\bar\nu_\mu, \nu_\tau$ and $\bar\nu_\tau$ and they have the advantage
that their flux is hardly modified once it leaves the sources.

Recently, the underground neutrino detectors at Kamiokande II and IMB
compiled neutrino flux between 100 MeV and 1.5 GeV and
concluded\Ref\eighteen{K.S. Hirata et al., Phys. Lett.
\underbar{B280}, 146 (1992).} that the ratio of $\nu_\mu$ and $\nu_e$ is 1
instead of 2.  The latter value of 2 is expected if the observed
neutrinos are produced in the atmosphere, since pions and kaons are
the neutrino source.  This riddle, called the muon neutrino deficiency
problem, is solved if neutrino oscillation $(\nu_\mu\to
\nu_x)$ ensued after production in the
atmosphere.\REFS\nineteen{E.W. Beier et al., Phys. Lett.
\underbar{B283}, 440 (1992).}\REFSCON\twenty{R. Becker-Szendy et al.,
Phys. Rev. Letters, to appear (1992).}\refsend
However, study of the up-going muon suggests that such oscillation
does not take place.\refmark{\twenty} Also, it can be shown that the
calculated atmospheric neutrino flux\Ref\twentyone{e.g., E.V. Bugaev
and V.A. Naumov, 20th International Cosmic Ray Conference, HE 4.1-18
(1987); Sov. J. Nucl. Phys. \underbar{45}, 857 (1987); G. Barr, T.K.
Gaisser and T. Stanev, Phys. Rev. \underbar{D39}, 3532 (1989).}tends to be
an overestimate.  From these considerations, it is very likely that
the neutrinos observed by the underground detectors are not
atmospheric but cosmic,
\ie\ most neutrinos observed are coming from outside the atmosphere of
the earth.  But, of course, an ordinary mechanism for neutrino
production ends up with $\nu_\mu/\nu_e \simeq 2$, by the same reason
as for atmospheric neutrinos.  The model proposed in this project is
the only one which predicts $\nu_\mu/\nu_e =1$.  Moreover, it is
worthwhile to mention that the neutrino flux spectrum inferred from
the underground detector data\refmark{\eighteen}is close to
$E^{-2.2}.$ Further observation of the neutrino flux will decide the
validity of the model.
\smallskip
\centerline{\fourteenrm\bf V.\ \  Summary and Further Predictions}

The prediction of universality for the energy spectrum from point
sources (with variation of the spectral index for individual
sources) is dramatically borne out by the GRO data and the cosmic
energy spectrum at the source.  Also an approximate equality of
$\nu_\mu$ and $\nu_e$ in the underground detector lends support to
this model.

In order to further confront the theory with observational data, I
propose to analyze the time variation of the $\gamma$-ray spectrum.
According to the model suggested in this article, the instantaneous
spectrum of $\gamma$-rays from point sources is Planckian.  Only after
integration over the various temperature, can one get a power law
spectrum as a reflection of a power law expansion rate.  For example,
the variability of X-rays from quasar 3C273 is $\sim$ 2 days.
Therefore, we need a slicing of the data in the time bin less than 2
days for 3C273.  Of course, the slicing of data into small time
intervals results in the data of poor statistics.  Therefore, an
appropriate integration of the data may be necessary.  In conclusion,
I propose to analyze the CGRO data and extract information on the time
variation of the spectrum.

The author is indebted to David Williams for reading the manuscript.
The work is supported in part by the U.S. Department of Energy.

\baselineskip=13pt
\refout

\end